\documentclass[doublecol]{epl2}

\newcommand{\eqref}[1]{(\ref{#1})}

\title{A discrete model of water with two distinct glassy phases}
\shorttitle{A discrete model of water}

\author{A. Pagnani\inst{1,2} \and M. Pretti\inst{3}}
\shortauthor{A. Pagnani and M. Pretti}

\institute{
  \inst{1} HuGeF Torino - Via Nizza 52, I-10126 Torino (Italy) \\
  \inst{2} CMP Politecnico di Torino - Corso Duca degli Abruzzi 24, I-10129 Torino (Italy) \\
  \inst{3} CNR-ISC and CNISM, Dip.to di Fisica, Politecnico di Torino - Corso Duca degli Abruzzi 24, I-10129 Torino (Italy)
}

\abstract{
We investigate a minimal model for non-crystalline water, defined
on a Husimi lattice. The peculiar random-regular nature of the
lattice is meant to account for the formation of a random
4-coordinated hydrogen-bond network. The model turns out to be
consistent with most thermodynamic anomalies observed in liquid
and supercooled-liquid water. Furthermore, the model exhibits two
glassy phases with different densities, which can coexist at a
first-order transition. The onset of a complex free-energy
landscape, characterized by an exponentially large number of
metastable minima, is pointed out by the cavity method, at the
level of 1-step replica symmetry breaking.
}

\begin{document}

\maketitle

\section{Introduction}

In the last decade, there have been several attempts at describing structural
glasses by means of simple lattice models, in the analytical framework of
random-regular (Bethe or Husimi)
lattices~\cite{Franz_et_al2001,BiroliMezard2002,Picaciamarra_et_al2003,WeigtHartmann2003,KrzakalaTarziaZdeborova2008}.
For these models, the cavity method generally predicts a discontinuous
replica-symmetry breaking, {\em i.e.}, a sudden emergence of an exponentially
large number of metastable free-energy minima.
On the other hand, great interest has been attracted by polyamorphism ({\em
i.e.}, the existence of different glass forms of the same
substance)~\cite{LoertingGiovambattista2006}, both because of the relative
novelty of the phenomenon (first discovered for water in
1985~\cite{MishimaCalvertWhalley1985}) and because of its relationship with the
popular ``second critical point'' conjecture, put forward by Stanley and
coworkers~\cite{PooleSciortinoEssmannStanley1992} to explain water
anomalies~\cite{Debenedetti2003}.

Lattice models have long been used for investigating
water~\cite{Bell1972,LavisSouthern1984,SastrySciortinoStanley1993,BesselingLyklema1994,RobertsDebenedetti1996,SastryDebenedettiSciortinoStanley1996,FranzeseStanley2002,FranzeseMarquesStanley2003,BuzanoPretti2004,HenriquesGuisoniBarbosaThieloBarbosa2005,HoyeLomba2010,BuzanoDestefanisPretti2008,PrettiBuzanoDestefanis2009}.
In some recent papers, coauthored by one of
us~\cite{BuzanoDestefanisPretti2008,PrettiBuzanoDestefanis2009}, it has been
shown that a first-order (quasi-chemical) approximation~\cite{LavisBell1999} on
a tetrahedral cluster is extremely effective to compute the phase diagram and
the thermodynamic properties of a special class of water-like models, derived
from the early Bell model~\cite{Bell1972}. These models, defined on the
(regular) body-centered cubic (bcc) lattice, do not contain a mechanism capable
of inducing glassy behavior, so that they exhibit only crystalline (ice-like)
phases at low temperature.

The present letter is motivated by the following observations. i) For a generic
lattice model, the aforementioned quasi-chemical approximation, with a given
choice of the associated cluster, coincides with the exact solution of a
corresponding model, defined on a Husimi lattice made up of clusters of the
same type~\cite{LavisBell1999,Pretti2003}, under the hypothesis of replica
symmetry~\cite{MezardParisi2001,MezardMontanari2006}. As a consequence, we can
define suitable Husimi lattice models, whose high-temperature behavior turns
out to be very similar to that of the original water-like bcc-lattice models.
ii) It is known from experiments that directional correlation of hydrogen bonds
in real water is almost completely lost after the second consecutive bond (see
\cite{CabaneVuilleumier2005} and references therein). The random nature of a
Husimi lattice can roughly describe a similar scenario (this was not possible
on a regular lattice). iii) We expect that the resulting frustration might
hamper the onset of ice-like order, allowing for the possibility of glassy
behavior at low temperature. Such a possibility can be easily investigated by
the cavity method~\cite{MezardParisi2001}.

Concerning point i), let us remark the nontrivial difference between a Husimi
{\em tree} (which is, a system with a boundary) and a Husimi {\em lattice}
(which is, a system without a boundary, in which all sites have the same
coordination number). The latter system locally exhibits the same tree-like
structure as the former, but contains loops on a larger
scale~\cite{MezardParisi2001}. The two systems may be considered equivalent as
long as the frustration arising from the presence of loops is not strong enough
to induce replica-symmetry breaking ({\em i.e.}, glassy behavior).

The main finding of this work is a {\em Husimi lattice} model predicting two
different glassy phases, which we are led to identify with the low- and
high-density amorphous (LDA, HDA) ices, observed
experimentally~\cite{LoertingGiovambattista2006,Debenedetti2003}. This model
provides interesting insights about the controversial nature of water
polyamorphism and, more in general, the metastable phase diagram of
water~\cite{Debenedetti2003}.

\section{The model}

\begin{figure}
  \onefigure{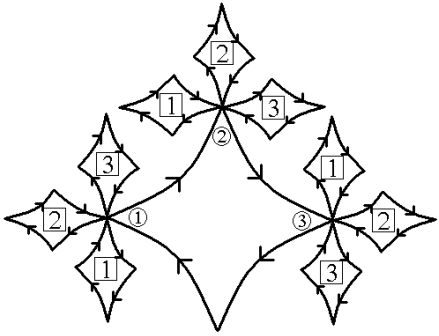}
  \caption
  {
    Portion of the Husimi lattice. Each site is connected to 4 plaquettes.
    The bottom site, connected only to the central plaquette,
    gives a pictorial view of the ``cavity bias'' or ``message'' concept (see the text).
  }
  \label{fig:husimi_lattice}
\end{figure}

We consider a Husimi lattice made up of {\em oriented} plaquettes of 4 sites,
as shown in fig.~\ref{fig:husimi_lattice}. Each site $i$ (characterized by a
configuration variable $x_i$) may be empty ($x_i=0$) or occupied by a water
molecule, in 2 possible configurations ($x_i=1,2$). Each water molecule
possesses four equivalent bonding arms, which can point toward nearest neighbor
sites, and may be either concordant ($x_i=1$) or discordant ($x_i=2$) with edge
orientations. An attractive potential energy $-\epsilon<0$ is assigned to any
pair of nearest-neighbor occupied sites. This is the ordinary van der Waals
contribution. A hydrogen bond is formed, yielding an extra energy $-\eta<0$,
whenever two nearest-neighbor molecules point an arm toward each other, with no
distinction between donors and acceptors (more precisely, a bond is formed on
an edge oriented from $i$ to $j$ iff $x_i=1$ and $x_j=2$). The hydrogen bond
may be weakened by the presence of extra molecules in the same plaquette,
namely, each extra molecule gives rise to an energy penalty $\eta c/2$, with
$c\in \left]0,1\right[$. The hamiltonian of the system can be written as a sum
over plaquettes
\begin{equation}
  \mathcal{H} = \sum_{\langle i,j,k,l \rangle} h_{x_i,x_j,x_k,x_l}
  \, ,
  \label{eq:hamiltonian}
\end{equation}
where site indices $i,j,k,l$ are enumerated according to the
plaquette orientation. The elementary contribution $h_{x,y,z,w}$
(``plaquette energy'') can be written as
\begin{equation}
  h_{x,y,z,w} = u_{x,y,z,w} + u_{y,z,w,x} + u_{z,w,x,y} + u_{w,x,y,z}
  \, ,
  \label{eq:plaquette_energy}
\end{equation}
where
\begin{equation}
  u_{x,y,z,w} = - \mu \frac{n_x}{4} - \epsilon n_x n_y
  - \eta b_{x,y} \left( 1 - c \frac{n_z+n_w}{2} \right)
  \, ,
  \label{eq:plaquette_energy_term}
\end{equation}
$n_x$ is an ``occupation function'' ($n_x=0$ if $x=0$; $n_x=1$
otherwise), $b_{x,y}$ is a ``bond function'' ($b_{x,y}=1$ if $x=1$
and $y=2$; $b_{x,y}=0$ otherwise), and $\mu$ is the chemical
potential (we study the grand-canonical ensemble). Looking at
eq.~\eqref{eq:plaquette_energy}, one immediately realizes that the
plaquette energy is invariant under circular permutations of the
configuration variables. As a consequence, the full hamiltonian
\eqref{eq:hamiltonian} is unaffected by the choice of the ``first
site'' $i$ in each plaquette.

For the sake of brevity, we have given only a formal description of the model.
Indeed, the latter is the Husimi-lattice version of a water-like model similar
to that proposed by Roberts and Debenedetti~\cite{RobertsDebenedetti1996}. This
can be easily deduced by comparing the original
paper~\cite{RobertsDebenedetti1996} with the tetrahedral cluster approximation
developed in~\cite{BuzanoPretti2004}. The square plaquettes of the Husimi
lattice correspond to tetrahedral clusters on the bcc lattice, whereas
plaquette orientations are related with the geometric structure of model
molecules. Let us remark that, even though the current model neglects some
details considered in \cite{RobertsDebenedetti1996} and \cite{BuzanoPretti2004}
(namely, the distinction between donors and
acceptors~\cite{RobertsDebenedetti1996}, and the presence of extra nonbonding
configurations~\cite{RobertsDebenedetti1996,BuzanoPretti2004}), it nonetheless
reproduces the typical thermodynamic anomalies observed in real water. Such a
result confirms that the physical mechanism underlying the anomalies is mainly
based on the ``weakening parameter'' $c$, which favors states characterized by
a positive correlation between higher local entropy (weaker bonds) and higher
local density~\cite{RobertsDebenedetti1996,BuzanoPretti2004}. A standard
statistical-mechanical argument~\cite{Debenedetti2003} relates such a positive
correlation to a negative thermal expansion coefficient, {\em i.e.}, to the
onset of a density maximum.

\section{The replica-symmetric (RS) solution}

According to the cavity method~\cite{MezardParisi2001}, in the RS
assumption, the model can be solved by a recursion equation for
so-called ``cavity biases'' or ``messages''. An elementary message
$m^{a \to i}_{x}$ represents the probability of the $x$
configuration at the $i$ site, when the latter is detached from
all plaquettes except $a$. With reference to
fig.~\ref{fig:husimi_lattice}, denoting by $\hat{m}_x$ the message
from the central plaquette to the bottom site, the recursion
equation reads
\begin{equation}
  \hat{m}_x = e^f \sum_{x_1=0}^{2} \sum_{x_2=0}^{2} \sum_{x_3=0}^{2}
  e^{-\beta h_{x,x_1,x_2,x_3}} \prod_{i=1}^{3} \prod_{a=1}^{3}
  m^{a \to i}_{x_i}
  \, ,
  \label{eq:recursion_equation}
\end{equation}
where $\beta=1/k_\mathrm{B}T$ is the inverse temperature and $f$ is a
normalization constant, ensuring that $\sum_{x=0}^{2}\hat{m}_x=1$. Let us note
that, in our case, since the lattice has no local heterogeneities, the messages
do not depend on the position (one can drop the $a \to i$ superscript).
Therefore, eq.~\eqref{eq:recursion_equation} eventually simplifies to a set of
three equations ($x=0,1,2$) of the fixed-point form
\begin{equation}
  m_x \propto \sum_{x_1=0}^{2} \sum_{x_2=0}^{2} \sum_{x_3=0}^{2}
  e^{-\beta h_{x,x_1,x_2,x_3}} \prod_{i=1}^{3} {m_{x_i}}^3
  \, ,
  \label{eq:recursion_equation_homogeneous}
\end{equation}
where we have omitted the trivial normalization constant. These equations can
be solved numerically by simple iteration. The actual probability $p_x$ of the
$x$ configuration can then be evaluated by considering the operation of
attaching four equivalent branches, like that depicted in
fig.~\ref{fig:husimi_lattice}, to the same ``root'' site~\cite{Pretti2003}. One
obtains $p_x \propto {m_x}^4$, where we have again omitted a normalization
constant, needed to ensure $\sum_{x=0}^{2}p_x=1$. The RS solution is always
characterized by $p_1=p_2$, {\em i.e.}, no preference between the two molecule
configurations. The mass density $\rho$ is simply related to the occupation
probability $p_\mathrm{occ}=1-p_0=p_1+p_2$ as
$\rho=(M/\upsilon)\,p_\mathrm{occ}$, where $M\approx18\,\mathrm{g/mol}$ is the
molecular mass of water and $\upsilon$ is the volume per site (which is an
adjustable parameter of the model).

The grand-canonical free energy per site $F/N$ (let $F$ and $N$ denote
respectively the thermodynamic free-energy times $\beta$, and the number of
lattice sites) can be evaluated as a function of the messages
as~\cite{Pretti2003}
\begin{equation}
  \frac{F}{N} = - \ln \frac{ \displaystyle
  \sum_{x_0=0}^{2} \sum_{x_1=0}^{2} \sum_{x_2=0}^{2} \sum_{x_3=0}^{2}
  e^{-\beta h_{x_0,x_1,x_2,x_3}} \prod_{i=0}^{3} {m_{x_i}}^3
  }{\displaystyle
  \left(\sum_{x=0}^{2} {m_x}^4\right)^3}
  \, .
  \label{eq:free_energy}
\end{equation}
Pressure can then be computed as $P=-(k_\mathrm{B}T/\upsilon)(F/N)$. In the
presence of multiple solutions of
eq.~\eqref{eq:recursion_equation_homogeneous}, {\em i.e.}, of competing phases,
the thermodynamically stable one is selected by the lowest free-energy (highest
pressure) value. In general, the knowledge of the cavity biases allows one to
compute all the thermodynamic properties of interest, including response
functions, such as the specific heat, the isothermal compressibility
$\kappa_T=(\partial\ln\rho/\partial P)_T$, and the isobaric thermal expansion
coefficient $\alpha_P=-(\partial\ln\rho/\partial T)_P$, whose anomalous
behavior is specially relevant in real water.
It is also possible to obtain numerically simple equations for the loci of
divergence of the response functions (spinodals) and the temperature of maximum
density (TMD) locus at constant pressure, defined by $\alpha_P=0$.

\begin{figure}
  \onefigure{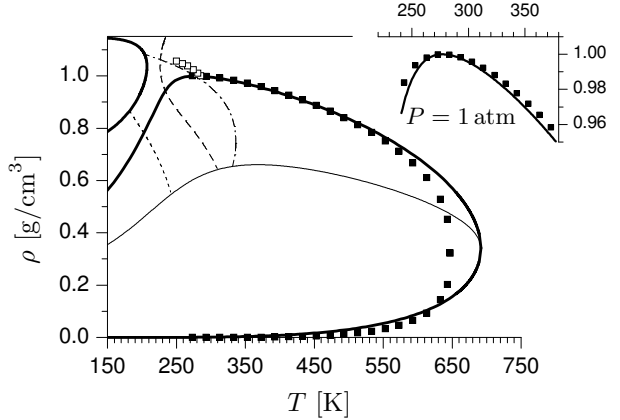}
  \caption
  {
    Temperature-density phase diagram in the RS assumption.
    Thick solid lines denote binodals.
    Thin lines are defined by the following conventions:
    A solid line denotes the liquid spinodal,
    a dash-dotted line denotes the TMD locus,
    a dashed line denotes the stability limit of the RS solution,
    and a dotted line denotes the zero-entropy locus.
    Solid and open squares denote experimental results for the
    liquid-vapor binodal~\cite{NIST} and the TMD locus~\cite{AngellKanno1976}
    (the latter data refer indeed to heavy water; pressure values have been
    converted to density values by \cite{NIST}).
    The inset displays the isobar at $P=1\un{atm}$,
    along with the experimental values~\cite{Chaplin} used for parameter fitting.
  }
  \label{fig:phase_diagram}
\end{figure}

A fitting procedure has been performed to fix the model parameters. The
hydrogen bond energy $\eta$ and the volume per site $\upsilon$ are easily
computed as a function of the van der Waals energy $\epsilon$ and the weakening
parameter $c$, by requiring the maximum density at $1\un{atm}$ to be
$1\un{g/cm^3}$ and to occur exactly at $3.984\un{^\circ C}$~\cite{Chaplin}. We
have then fitted only the two parameters $\epsilon$ and $c$ by minimizing (in a
least-square sense)\footnote{We have used the MATLAB routine {\tt lsqnonlin}.}
the difference between a set of experimental density values at $1\un{atm}$
(fig.~\ref{fig:phase_diagram}, inset) and the corresponding theoretical values.
The final result is: $\eta\approx7.786\un{kJ/mol}$,
$\epsilon\approx3.621\un{kJ/mol}$, $c\approx0.6450$,
$\upsilon\approx15.70\un{cm^3/mol}$.

In fig.~\ref{fig:phase_diagram} we report the RS phase diagram in the
temperature-density plane. We obtain a quite good agreement with experimental
data for the whole liquid-vapor binodal curve (not fitted). The critical
exponent is not correct, due to the mean-field nature of the model, but the
critical density $\rho_\mathrm{c}\approx0.3417\un{g/cm^3}$ is remarkably close
to the experimental value $0.322\un{g/cm^3}$~\cite{Chaplin}. The TMD locus
correctly displays a negative slope, quantitatively similar to the experimental
one. Moreover, in the metastable region of the liquid phase at very low
densities (negative pressures), the TMD locus exhibits a slight reentrance
({\em i.e.}, positive slope), which has been predicted by simulations with
various intermolecular
potentials~\cite{NetzStarrStanleyBarbosa2001,YamadaMossaStanleySciortino2002}.
At odd with such simulations, our TMD line eventually meets the spinodal line.
In the pressure-temperature diagram, the two curves meet tangentially at a
pressure minimum, as required for thermodynamic
consistency~\cite{PooleSciortinoEssmannStanley1993}, but the spinodal is not
found to re-enter the positive pressure region. The latter fact contradicts
Speedy's early conjecture~\cite{Speedy1982} about the divergent-like behavior
of response functions in the supercooled liquid regime. Conversely, our model
seems to support Stanley's conjecture, as it predicts, at very low temperature,
a ``second critical point'', terminating a coexistence region of two different
liquid-like phases. Unfortunately, all this coexistence region lies below the
stability limit of the RS solution (dashed line in
fig.~\ref{fig:phase_diagram}), where the latter is no longer valid. The
inadequacy of such a solution in this regime is also pointed out by the fact
that its entropy becomes negative below a given temperature (dotted line in
fig.~\ref{fig:phase_diagram}).

\begin{figure}
  \onefigure{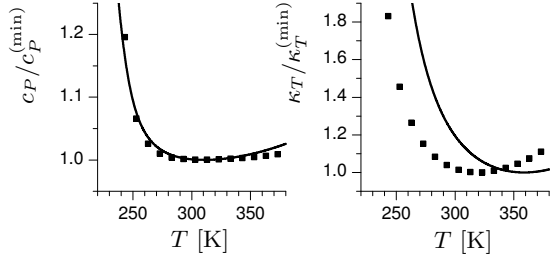}
  \caption
  {
    Normalized isobaric specific heat (left) and isothermal compressibility (right)
    as a function of temperature at $P=1\un{atm}$.
    Lines and symbols denote respectively theoretical and experimental~\cite{Chaplin} results.
    The theoretical (experimental) minima occur at $37.45\un{^\circ C}$
    ($36\un{^\circ C}$)
    and $85.58\un{^\circ C}$ ($46.5\un{^\circ C}$).
  }
  \label{fig:response_functions}
\end{figure}
Fig.~\ref{fig:response_functions} displays the isobaric specific
heat and the isothermal compressibility as a function of
temperature, at atmospheric pressure. The well-known anomalous
behavior of these response functions, related respectively to
entropy and density fluctuations~\cite{Debenedetti2003}, is
reproduced by the model in a qualitatively correct fashion. This
result is achieved without the singularity required by Speedy's
conjecture. Indeed, upon decreasing temperature, the theoretical
curves do not exhibit any divergence, but only pronounced maxima.
Quantitative agreement with experiments is not so good in this
case, so that we report data normalized to the minimum values
$c_P^\mathrm{(min)}$ and $\kappa_T^\mathrm{(min)}$.

\section{The glassy phases}

Let us now investigate what happens when the RS ansatz does not hold. We make
use of the cavity method at the level equivalent to 1-step replica symmetry
breaking (RSB)~\cite{MezardParisi2001}. For a given free-energy landscape, the
complexity function $\Sigma(F)$ is defined as the log-number of minima (``pure
states'', or simply ``states'') with free energy in the range $[F,F+\upd F]$.
If the number of states is exponentially large, then $\Sigma$ is an extensive
quantity. One can define a pseudo free-energy (``replicated'' free-energy)
as~\cite{Monasson1995}
\begin{equation}
  \Phi(\psi) = - \ln \sum_\alpha e^{- \psi F_\alpha}
  \, ,
  \label{eq:replicated_free_energy}
\end{equation}
where $F_\alpha$ is the free energy of the
$\alpha$ state, the sum runs over all states, and $\psi$ is a
pseudo inverse-temperature (Parisi parameter). A saddle-point
calculation shows that $\Phi(\psi)$ is the Legendre transform of
$\Sigma(F)$, so that $\Phi(\psi)$ allows one to reconstruct
$\Sigma(F)$ via the parametric representation
\begin{equation}
  F(\psi) = \frac{\upd\Phi}{\upd\psi}(\psi)
  \, , \qquad
  \Sigma(\psi) = \psi \frac{\upd\Phi}{\upd\psi}(\psi) - \Phi(\psi)
  \, .
\end{equation}
The replicated free-energy (and all the thermodynamic quantities) can be
computed, as a function of $\psi$, by solving an integral equation for the
message distribution\footnote{Hereafter, $m$ without subscript denotes a
normalized 3-component array $(m_0,m_1,m_2)$.} $\mathcal{P}(m)$ over the states
\begin{equation}
  \mathcal{P}(m) = e^\phi \int \delta(m-\hat{m}) \, e^{- \psi f} \prod_{i=1}^{3} \prod_{a=1}^{3} \mathcal{P}(m^{a \to i}) \, \upd m^{a \to i}
  \, ,
  \label{eq:integral_equation}
\end{equation}
where both $\hat{m}$ and $f$ are defined by eq.~\eqref{eq:recursion_equation}
as functions of the set of incoming messages $\{m^{a \to i}\}$, and $\phi$
ensures the normalization condition $\int\mathcal{P}(m)\,\upd m=1$.
Equation~\eqref{eq:integral_equation} can be solved numerically by a population
dynamics technique~\cite{MezardParisi2001}.


In the RS phase, $\mathcal{P}(m)$ is a delta-function and
eq.~\eqref{eq:recursion_equation} is recovered. The stability limit of the RS
phase, displayed in fig.~\ref{fig:phase_diagram}, can be determined by assuming
that $\mathcal{P}(m)$ has only a slight variance around the RS value. One
obtains a linearized ``propagation equation'' for the covariance matrix of
$\mathcal{P}(m)$
\begin{equation}
  \langle \delta m_x \delta m_y \rangle = \sum_{x'=0}^{2} \sum_{y'=0}^{2} K_{x,y;x',y'} \langle \delta m_{x'} \delta m_{y'} \rangle
  \, ,
\end{equation}
where the ($9 \times 9$) ``transfer matrix'' is
\begin{equation}
  K_{x,y;x',y'} = \sum_{i=1}^{3} \sum_{a=1}^{3}
  \frac{\partial \hat{m}_x}{\partial m_{x'}^{a \to i}}
  \frac{\partial \hat{m}_y}{\partial m_{y'}^{a \to i}}
  \, ,
  \label{eq:transfer_matrix}
\end{equation}
and the Jacobians $\partial \hat{m}_x / \partial m_{x'}^{a \to i}$
can be evaluated from eq.~\eqref{eq:recursion_equation}. The RS
solution becomes unstable when the maximum eigenvalue of the
transfer matrix \eqref{eq:transfer_matrix} becomes larger than 1.

\begin{figure}
  \onefigure{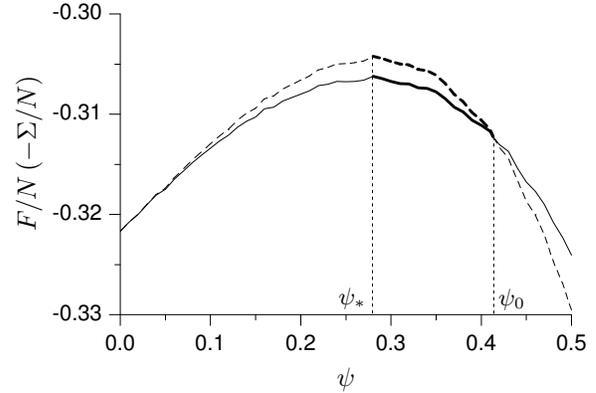}
  \caption
  {
    $F/N$ (dashed line) and $(F-\Sigma)/N$ (solid line)
    as a function of $\psi$
    at $T=135\un{K}$, $\mu/\eta=-2.38$.
    Thinner lines denote unphysical parts of the curves.
  }
  \label{fig:cavity_rsb_1}
\end{figure}

\begin{figure}
  \onefigure{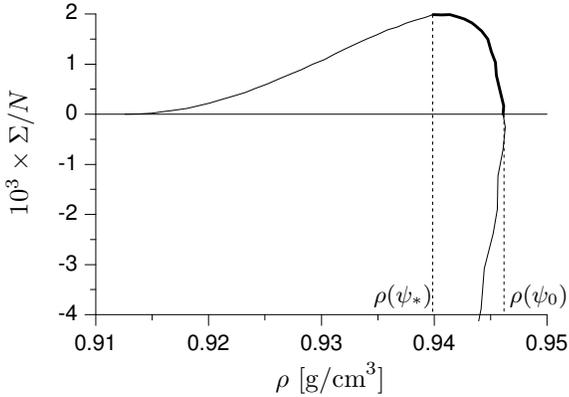}
  \caption
  {
    Parametric plot of $\Sigma(\psi)/N$ {\em vs.} $\rho(\psi)$
    at $T=135\un{K}$, $\mu/\eta=-2.38$
    (as in fig.~\ref{fig:cavity_rsb_1}).
    Thinner lines denote unphysical parts of the curve.
  }
  \label{fig:cavity_rsb_2}
\end{figure}

In the RSB phase, the thermodynamically relevant states are
identified by a minimum (wrt $\psi$) of the function $F-\Sigma$,
which properly takes into account that the probability measure may
be split over a large number of states with free energy around
$F$. We report the typical behavior of $F/N$ and $(F-\Sigma)/N$ in fig.~\ref{fig:cavity_rsb_1}.
The leftmost part of the curves ($\psi<\psi_*$) is unphysical,
because $\upd F/\upd\psi=\upd^2\Phi/\upd\psi^2<0$, which is inconsistent with
eq.~\eqref{eq:replicated_free_energy}. The rightmost part
($\psi>\psi_0$) is also unphysical, because $\Sigma<0$. Therefore,
the physical minimum of $F-\Sigma$ is attained at the crossing
point $\psi_0$, such that $\Sigma(\psi_0)=0$. In the jargon of
replica theory, this is called a ``condensed'' glass phase. The
number of thermodynamically relevant states is sub-exponential,
but there exists an exponentially large number of metastable
states. Fig.~\ref{fig:cavity_rsb_2} shows that the former
states exhibit the largest density, $\rho(\psi_0)$, whereas the latter
span a range of smaller density values,
$\rho(\psi_*)<\rho<\rho(\psi_0)$.

Upon increasing temperature, we observe that the RSB phase undergoes a {\em
continuous} transition to the RS phase, at odd with the glass models cited
above~\cite{Franz_et_al2001,BiroliMezard2002,Picaciamarra_et_al2003,WeigtHartmann2003,KrzakalaTarziaZdeborova2008}.
As previously mentioned, those models predict (upon {\em decreasing}
temperature) an abrupt appearance of many metastable states (``dynamical
transition''), where the system is expected to get stuck during quenching. Such
a difference is rather intriguing. In real water, the glass transition appears
to be much ``weaker'' than that found in the usual molecular
liquids~\cite{Angell2008}, and the very existence of a glass transition in
water (before recrystallization upon heating) has been much
debated~\cite{YueAngell2004,Kohl_et_al2005}. This ``weakness'' is believed to
reflect a greater ease for the system to explore its energy landscape, {\em
i.e.}, a greater accessibility of the ideal glass state~\cite{Angell2008}.
Qualitatively speaking, such a scenario might be consistent with the absence of
a dynamical transition, as predicted by our model.

\begin{figure}
  \onefigure{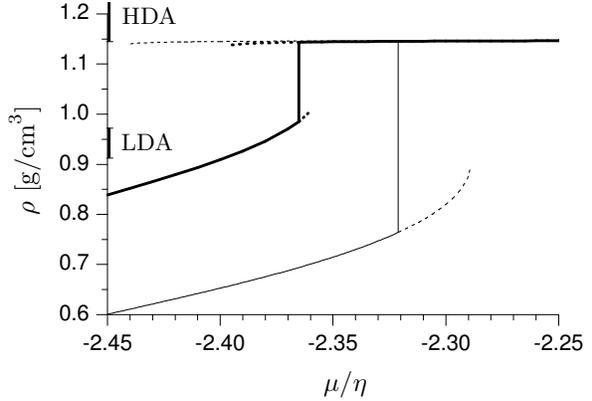}
  \caption
  {
    Density $\rho$ as a function of the normalized chemical
    potential $\mu/\eta$ at $T=135\un{K}$.
    Thinner lines denote the (unstable) RS solution;
    thicker lines denote $\rho(\psi_0)$ ({\em i.e.}, the density of the
    thermodinamically relevant states) in the RSB solution.
    Dotted lines denote metastable continuations.
    Experimental density ranges of LDA and HDA ices~\cite{Mishima1994}
    are highlighted on the left axis.
  }
  \label{fig:lda_hda_1}
\end{figure}
Let us also consider the RSB phase at constant temperature.
Fig.~\ref{fig:lda_hda_1} shows that, upon increasing the chemical potential, a
higher density solution appears discontinuously. It turns out that even the
latter is a condensed glass phase, in the previously explained sense. As
previously mentioned, we find it natural to identify these phases with the LDA
and HDA ices. Remarkably, the density range predicted by the RSB solution for
LDA is consistent with experimental values, at odd with the low-density
(unstable) RS solution. The behavior of HDA, as a function of the Parisi
parameter $\psi$, is qualitatively similar to that of LDA, displayed in
fig.~\ref{fig:cavity_rsb_1}. Nevertheless, the density values of the metastable
states are {\em higher} than the density of the thermodynamically relevant
states, {\em i.e.}, $\rho(\psi_*)>\rho(\psi_0)$ (the $\rho$-$\Sigma$ plot is a
mirror image of that of fig.~\ref{fig:cavity_rsb_2}). Such a result guarantees
that, even in the region where both solutions exist, they are always separated
by a density gap with no states. As a consequence, our model supports the
picture of a real first-order-like LDA-HDA
transition~\cite{LoertingGiovambattista2006,Debenedetti2003}, excluding the
possibility of a smooth crossover, suggested by some experimental
observations~\cite{Tulk_et_al2002}. The transition pressure at $T=135\un{K}$
turns out to be about $28.6\un{MPa}$, much lower than the experimental estimate
($\sim 200\un{MPa}$)~\cite{Mishima1994}, but the decreasing trend, upon
increasing temperature, is correctly reproduced.

\begin{figure}
  \onefigure{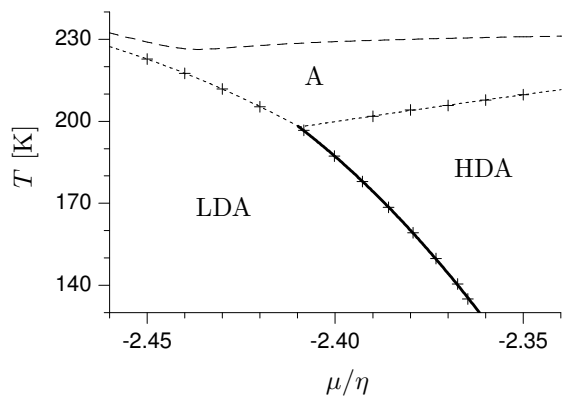}
  \caption
  {
    RSB phase diagram (see the text).
    The dashed line denotes the RS stability limit. Other
    lines are 2nd order polynomial interpolations of symbols.
  }
  \label{fig:lda_hda_2}
\end{figure}
The transition line terminates in a critical point, which occurs around
$T\approx198\un{K}$, $P\approx16.7\un{MPa}$, and above which the two glassy
phases become indistinguishable. Note that, since LDA and HDA are,
respectively, characterized by $\rho(\psi_*)<\rho(\psi_0)$ and
$\rho(\psi_*)>\rho(\psi_0)$, indistinguishability implies
$\rho(\psi_*)=\rho(\psi_0)$ in the supercritical region. This fact gives rise
to a third kind of RSB phase (A), characterized by $\psi_0=\psi_*=0$ (see
footnote\footnote{The corresponding complexity is $\Sigma=0$; $\psi$ values
other than $0$ result in a negative complexity.}). The maximal complexity
$\Sigma(\psi_*)$ of LDA and HDA turns out to vanish in a continuous way,
defining two more (LDA-A and HDA-A) transition lines. Fig.~\ref{fig:lda_hda_2}
shows numerical evidence that such lines meet precisely at the critical point.
The A-phase may still be considered a glassy phase, since the message
distribution $\mathcal{P}(m)$ is still nontrivial. As previously mentioned, the
vanishing complexity means that such a distribution represents a
sub-exponential number of pure states. As shown in fig.~\ref{fig:lda_hda_2},
the A-phase region extends until the RS stability limit, where $\mathcal{P}(m)$
eventually degenerates into a Dirac delta, and the RS solution becomes stable.

\section{Summary and conclusions}

We have studied a simplified model of water, defined on a Husimi lattice, which
can be solved by the cavity method. Despite its simplicity, the model seems to
capture a lot of features arising in real (and simulated) water, including
thermodynamic anomalies of the liquid phase and a low-temperature glassy phase
displaying polyamorphism. To the best of our knowledge, this is the first
lattice model describing all these features together.

The glassy behavior originates in the competition between the tendency of the
model to form regular hydrogen-bond networks, driven by the energy term $-\eta
b_{x,y}$ of eq.~\eqref{eq:plaquette_energy_term}, and the topological disorder
of the Husimi lattice, which induces frustration. In other words, the lattice
randomness plays a relevant role in the model behavior, as it is responsible
for suppressing the ice-like phases that appear in the corresponding regular
lattice models. Even though the latter feature might be considered a weakness
of the model, we believe that it roughly illustrates the physical mechanism
underlying glassy behavior in a hydrogen-bonded (network-forming) fluid. In
this framework, the onset of polyamorphism is related to the fact that, due to
the weakening term $+\eta b_{x,y} c (n_z+n_w)/2$, the original model predicts a
low density network structure (stable at low pressure), besides the
``ordinary'' crystal-like ground state. The glass-glass transition may thus be
viewed as a ``frustrated'' reminiscence of the coexistence between these two
phases.

In the future, it might be interesting to investigate whether a similar
mechanism could give rise to glass-glass transitions even for lattice models
predicting a dynamical transition
scenario~\cite{Franz_et_al2001,BiroliMezard2002,Picaciamarra_et_al2003,WeigtHartmann2003,KrzakalaTarziaZdeborova2008}.
A more technical issue, still deserving investigation, is the stability of the
current picture with respect to further RSB steps.


\begin{thebibliography}{0}

\bibitem{Franz_et_al2001}
  \Name{Franz~S., M\'ezard~M., Ricci-Tersenghi~F., Weigt~M. \and Zecchina~R.}
  \REVIEW{Europhys. Lett.}{55}{2001}{465}.
\bibitem{BiroliMezard2002}
  \Name{Biroli~G. \and M\'ezard~M.}
  \REVIEW{Phys. Rev. Lett.}{88}{2002}{025501}.
\bibitem{Picaciamarra_et_al2003}
  \Name{Pica~Ciamarra~M., Tarzia~M., de~Candia~A. \and Coniglio~A.}
  \REVIEW{Phys. Rev. E}{67}{2003}{057105}.
\bibitem{WeigtHartmann2003}
  \Name{Weigt~M. \and Hartmann~A.~K.}
  \REVIEW{Europhys. Lett.}{62}{2003}{533}.

\bibitem{KrzakalaTarziaZdeborova2008}
  \Name{Krzakala~F., Tarzia~M. \and Zdeborov\'a~L.}
  \REVIEW{Phys. Rev. Lett.}{101}{2008}{165702}.

\bibitem{LoertingGiovambattista2006}
  \Name{Loerting~T. \and Giovambattista~N.}
  \REVIEW{J. Phys.: Condens. Matter}{18}{2006}{R919}.

\bibitem{MishimaCalvertWhalley1985}
  \Name{Mishima~O., Calvert~L.~D. \and Whalley~E.}
  \REVIEW{Nature}{314}{1985}{76}.

\bibitem{PooleSciortinoEssmannStanley1992}
  \Name{Poole~P.~H., Sciortino~F., Essmann~U. \and Stanley~H.~E.}
  \REVIEW{Nature}{360}{1992}{324}.

\bibitem{Debenedetti2003}
  \Name{Debenedetti~P.~G.}
  \REVIEW{J. Phys.: Condens. Matter}{15}{2003}{R1669}.

\bibitem{Bell1972}
  \Name{Bell~G.~M.}
  \REVIEW{J. Phys. C}{5}{1972}{889}.

\bibitem{LavisSouthern1984}
  \Name{Lavis~D.~A. \and Southern~B.~W.}
  \REVIEW{J. Stat. Phys.}{35}{1984}{489}.

\bibitem{SastrySciortinoStanley1993}
  \Name{Sastry~S., Sciortino~F., \and Stanley~H.~E.}
  \REVIEW{J. Chem. Phys.}{98}{1993}{9863}.

\bibitem{BesselingLyklema1994}
  \Name{Besseling~N.~A.~M. \and Lyklema~J.}
  \REVIEW{J. Phys. Chem.}{98}{1994}{11610}.

\bibitem{RobertsDebenedetti1996}
  \Name{Roberts~C.~J. \and Debenedetti~P.~G.}
  \REVIEW{J. Chem. Phys.}{105}{1996}{658}.

\bibitem{SastryDebenedettiSciortinoStanley1996}
  \Name{Sastry~S., Debenedetti~P.~G., Sciortino~F. \and Stanley~H.~E.}
  \REVIEW{Phys. Rev. E}{53}{1996}{6144}.

\bibitem{FranzeseStanley2002}
  \Name{Franzese~G. \and Stanley~H.~E.}
  \REVIEW{J. Phys.: Condens. Matter}{14}{2002}{2201}.

\bibitem{FranzeseMarquesStanley2003}
  \Name{Franzese~G., Marqu\'es~M.~I. \and Stanley~H.~E.}
  \REVIEW{Phys. Rev. E}{67}{2003}{011103}.

\bibitem{BuzanoPretti2004}
  \Name{Buzano~C. \and Pretti~M.}
  \REVIEW{J. Chem. Phys.}{121}{2004}{11856}.

\bibitem{HenriquesGuisoniBarbosaThieloBarbosa2005}
  \Name{Henriques~V.~B., Guisoni~N., Barbosa~M.~A.~A., Thielo~M. \and Barbosa~M.~C.}
  \REVIEW{Mol. Phys.}{103}{2005}{3001}.

\bibitem{HoyeLomba2010}
  \Name{H{\o}ye~J.~S. \and Lomba~E.}
  \REVIEW{Mol. Phys.}{108}{2010}{51}.

\bibitem{BuzanoDestefanisPretti2008}
  \Name{Buzano~C., De~Stefanis~E. \and Pretti~M.}
  \REVIEW{J. Chem. Phys.}{129}{2008}{024506}.

\bibitem{PrettiBuzanoDestefanis2009}
  \Name{Pretti~M., Buzano~C. \and De~Stefanis~E.}
  \REVIEW{J. Chem. Phys.}{131}{2009}{224508}.

\bibitem{LavisBell1999}
  \Name{Lavis~D.~A. \and Bell~G.~M.}
  \Book{Statistical Mechanics of Lattice Systems}
  \Vol{1}
  \Publ{Springer, Berlin}
  \Year{1999}
  \Page{173}.

\bibitem{Pretti2003}
  \Name{Pretti~M.}
  \REVIEW{J. Stat. Phys.}{111}{2003}{993}.

\bibitem{MezardParisi2001}
  \Name{M\'ezard~M. \and Parisi~G.}
  \REVIEW{Eur. Phys. J. B}{20}{2001}{217}.

\bibitem{MezardMontanari2006}
  \Name{M\'ezard~M. \and Montanari~A.}
  \REVIEW{J. Stat. Phys.}{124}{2006}{1317}.

\bibitem{CabaneVuilleumier2005}
  \Name{Cabane~B. \and Vuilleumier~R.}
  \REVIEW{C. R. Geosci.}{337}{2005}{159}.

\bibitem{Chaplin}
  \Name{Chaplin M.}
  {\tt http://www1.lsbu.ac.uk/water/}.

\bibitem{NIST}
  \Editor{Linstrom~P.~J. \and Mallard~W.~G.}
  \Book{NIST Chemistry WebBook, NIST Standard Reference Database Number 69}
  \Publ{National Institute of Standards and Technology, Gaithersburg MD, 20899}
  {\tt http://webbook.nist.gov/},
  \Year{retrieved November 21, 2008}.

\bibitem{AngellKanno1976}
  \Name{Angell~C.~A. \and Kanno~H.}
  \REVIEW{Science}{193}{1976}{1121}.

\bibitem{NetzStarrStanleyBarbosa2001}
  \Name{Netz~P.~A., Starr~F.~W., Stanley~H.~E. \and Barbosa~M.~C.}
  \REVIEW{J. Chem. Phys.}{115}{2001}{344}.
\bibitem{YamadaMossaStanleySciortino2002}
  \Name{Yamada~M., Mossa~S., Stanley~H.~E. \and Sciortino~F.}
  \REVIEW{Phys. Rev. Lett.}{88}{2002}{195701}.

\bibitem{PooleSciortinoEssmannStanley1993}
  \Name{Poole~P.~H., Sciortino~F., Essmann~U. \and Stanley~H.~E.}
  \REVIEW{Phys. Rev. E}{48}{1993}{3799}.

\bibitem{Speedy1982}
  \Name{Speedy~R.~J.}
  \REVIEW{J. Phys. Chem.}{86}{1982}{982}.

\bibitem{Monasson1995}
  \Name{Monasson~R.}
  \REVIEW{Phys. Rev. Lett.}{75}{1995}{2847}.

\bibitem{Angell2008}
  \Name{Angell~C.~A.}
  \REVIEW{Science}{319}{2008}{582}.

\bibitem{YueAngell2004}
  \Name{Yue~Y. \and Angell~C.~A.}
  \REVIEW{Nature}{427}{2004}{717};
  \SAME{435}{2005}{E1}.
\bibitem{Kohl_et_al2005}
  \Name{Kohl~I., Bachmann~L., Mayer~E., Hallbrucker~A. \and Loerting~T.}
  \REVIEW{Nature}{435}{2005}{E1}.

\bibitem{Tulk_et_al2002}
  \Name{Tulk~C.~A. {\em et al.}}
  \REVIEW{Science}{297}{2002}{1320}.

\bibitem{Mishima1994}
  \Name{Mishima~O.}
  \REVIEW{J. Chem. Phys.}{100}{1994}{5910}.

\end{thebibliography}
\end{document}